\documentclass[10pt,letterpaper,twocolumn,english,aps,showpacs]{revtex4}
\usepackage[T1]{fontenc}
\usepackage[latin9]{inputenc}
\usepackage{amstext}
\usepackage{amssymb}
\usepackage[pdftex]{graphicx}
\usepackage{amsmath}
\usepackage{epstopdf}
\usepackage{grffile}
\usepackage{todo}

\usepackage[colorlinks]{hyperref}

\makeatletter

\usepackage{babel}
\begin{document}

\title{Dynamical phase diagram of Gaussian wave packets in optical lattices}

\author{H. Hennig$^{1,2}$, T. Neff$^{1}$, R. Fleischmann$^{1}$}

\affiliation{$^{1}$Max Planck Institute for Dynamics and Self-Organization, 37073 G\"ottingen, Germany}
\affiliation{$^{2}$Department of Physics, Harvard University, Cambridge, MA 02138, USA}

\date{\today}
\begin{abstract}
We study the dynamics of self-trapping in Bose-Einstein condensates (BECs) loaded in deep optical lattices with Gaussian initial conditions, when the dynamics is well described by the Discrete Nonlinear Schr\"odinger Equation (DNLS). In the literature an approximate dynamical phase diagram based on a variational approach was introduced to distinguish different dynamical regimes: diffusion, self-trapping and moving breathers. However, we find that the actual DNLS dynamics shows a completely different diagram than the variational prediction. We numerically calculate a detailed dynamical phase diagram accurately describing the different dynamical regimes.
It exhibits a complex structure which can readily be tested in current experiments in BECs in optical lattices and in optical waveguide arrays. Moreover, we derive an explicit theoretical estimate for the transition to self-trapping in excellent agreement with our numerical findings, which may be a valuable guide as well for future studies on a quantum dynamical phase diagram based on the Bose-Hubbard Hamiltonian.
 
\end{abstract}

\pacs{03.75.Gg, 03.75.Lm, 67.85.Hj}

\maketitle
Bose-Einstein condensates (BECs) trapped in periodic optical potentials have proven to be an invaluable tool to study  fundamental and applied aspects of  quantum optics, quantum computing and solid state physics~\cite{Bloch:2005uv,Bakr:2009bx,Bakr:2010gd,Simon:2011ep,Morsch:2006em}.
In the limit of large atom numbers per well, the dynamics can be well described by a mean-field approximation which leads to a lattice version of the Gross-Pitaevskii equation,  the discrete nonlinear Schr\"odinger equation (DNLS) ~\cite{PETHICK:2008tn,Morsch:2006em}. One of the most intriguing features of the dynamics of nonlinear lattices is that excitations can spontaneously stably localize even for repulsive nonlinearities. This phenomenon of discrete self-trapping, also referred to as the formation of \textit{discrete breathers (DB)}, 
is a milestone discovery in nonlinear science that has sparked many studies (for reviews see~\cite{Flach:2008ud,Campbell:2004vz, Hennig:2013ja}). DBs have been observed experimentally in various physical systems such as arrays of nonlinear waveguides \cite{Eisenberg:1998wp,Morandotti:1999tt} and Josephson junctions \cite{Trias:2000uw,Ustinov:2003uc}, spins in antiferromagnetic solids \cite{Schwarz:1999vb,Sato:2004vw}, and BECs in optical lattices \cite{Eiermann:2004vt}. 
Self-trapping has also been shown to exist in the dynamics of the Bose-Hubbard Hamiltonian \cite{Hennig:2012gn} and in calculations beyond the Bose-Hubbard model, which include higher-lying states in the individual wells \cite{Sakmann:2009wy,TrujilloMartinez:2009gh}.

 Under which conditions will a Gaussian distributed initial condition in an optical lattice become diffusive, or self-trapped or a moving breather after sufficient propagation time?
This question was addressed in a seminal work \cite{Trombettoni:2001wl} that has become a standard reference in both experimental and theoretical studies involving self-trapping in optical lattices including BECs and optical waveguide arrays in the past decade \cite{Morsch:2006em, Christodoulides:2003vu, Fleischer:2003gc, Albiez:2005wk, Cataliotti:2001tx, Fleischer:2003cr, Eiermann:2004vt, Morsch:2001hj, Burger:1999vs, Smerzi:2002tj, Anker:2005vf, Iwanow:2004df, Eiermann:2003ja, Reinhard:2013ft, Bersch:2012ih, Dienst:2013gk, Dong:2013ky}. By means of a variational approach that approximates the DNLS dynamics it was shown  that the dynamical phase diagram is divided in in different regimes (diffusion, self-trapping and moving breathers) \cite{Trombettoni:2001wl}. The `dynamical phase diagram' distinguishes between qualitatively different steady state solutions. 
However, a recent study~\cite{Franzosi:2011ft} indicated that numerical simulations based on the DNLS show strong deviations from the variational dynamics. Moreover, as we show in Fig.~\ref{fig:colorportrait_alpha1}a, the actual parameter regions in which the different dynamical phases can be observed are completely different from those predicted in ref.~\cite{Trombettoni:2001wl}. Therefore, a new theory of the self-trapping of Gaussian wave packets is needed that can predict the dynamical regimes. 

Here, we study Gaussian wave packets of BECs in deep optical lattices in the mean-field limit (described by the DNLS). We both analytically and numerically present a detailed and accurate dynamical phase diagram that separates the different dynamical regimes (diffusion, self- trapping, moving breathers and multi-breathers). 

Although our focus is on BECs in optical lattices, where our dynamical phase diagram can be readily probed with single-site addressability \cite{Gericke:2008tq, Albiez:2005wk,Morsch:2006em}, our results apply as well other systems described by the DNLS including optical waveguide arrays~\cite{Christodoulides:2003vu,Christodoulides:2001tm}.
Our results may as well be a valuable guide for studies that aim to understand the deviations of the correlated self-trapping dynamics as described by the Bose-Hubbard Hamiltonian from the mean field dynamics. Examples of such deviations have been observed both experimentally \cite{Reinhard:2013ft} and theoretically \cite{Hennig:2012gn}.

\begin{figure}[t!]
  \centering
     \includegraphics{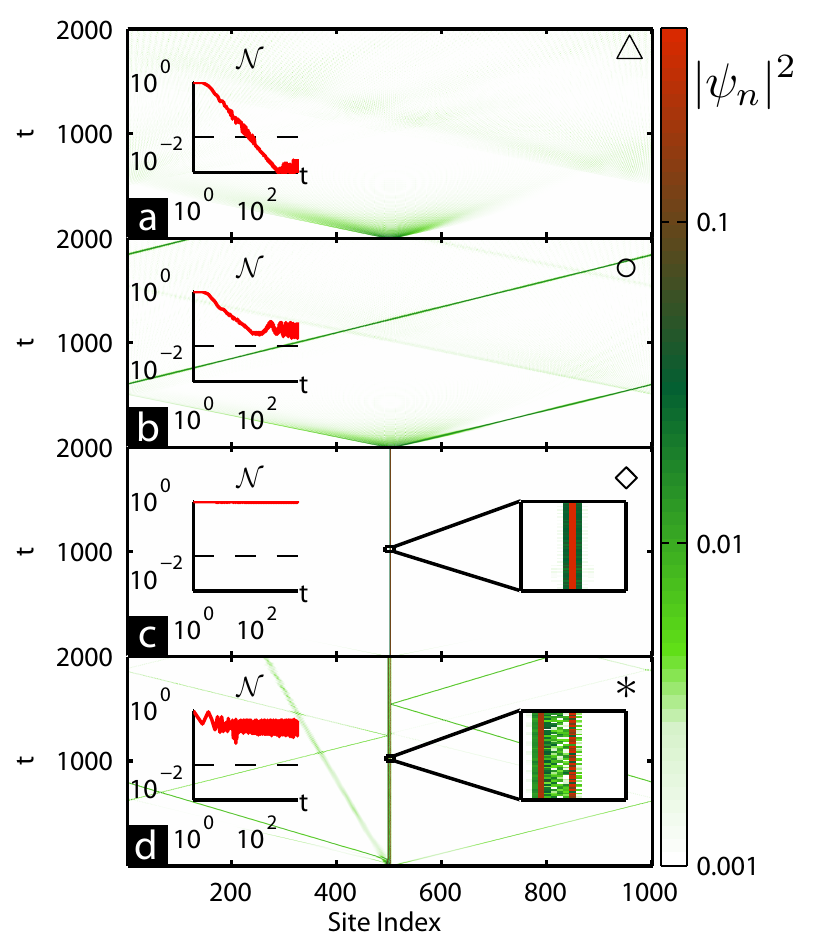}
     \caption{Examples of the different dynamical regimes and comparison with the variational approach \cite{Trombettoni:2001wl}. The density plots show the evolution of the norms $|\psi_n|^2$ of initially Gaussian wave packets (Eq.~\ref{ic}). The insets on the left show the time traces of the maximum local norm $\mathcal{N}$ (Eq.~\ref{eq:NJmax}).
     \textbf{(a)} Strongly diffusive regime. In contrast, the variational approach predicts self-trapping. Parameters: $\alpha_0=1$, $\cos(p_0)=0.88$, $\lambda=2.5$. 
     \textbf{(b)} Moving breather regime. The variational approach predicts diffusion. Parameters: $\alpha_0=1$, $\cos(p_0)=0.88$, $\lambda=1.5$.  
     \textbf{(c)} ($\alpha_0=1$, $\cos(p_0)=-1$, $\lambda=3$) DB solution.
     \textbf{(d)} ($\alpha_0=4$, $\cos(p_0)=-0.6$, $\lambda=8.9$) shows a breather of higher order with asymmetric shape. It corresponds to a drop in $\mathcal{H}_{thrs}$.
     The examples are indicated by $\bigtriangleup$ (a), $\bigcirc$ (b), $\Diamond$ (c) in Fig.(\ref{fig:colorportrait_alpha1}) (and $*$ (d) in Fig.(\ref{fig:colorportrait_alpha4}) in the Supplemental Material). }
     \label{fig:dens00}
 \end{figure}

\begin{figure*}[t!]
  \centering
      \includegraphics{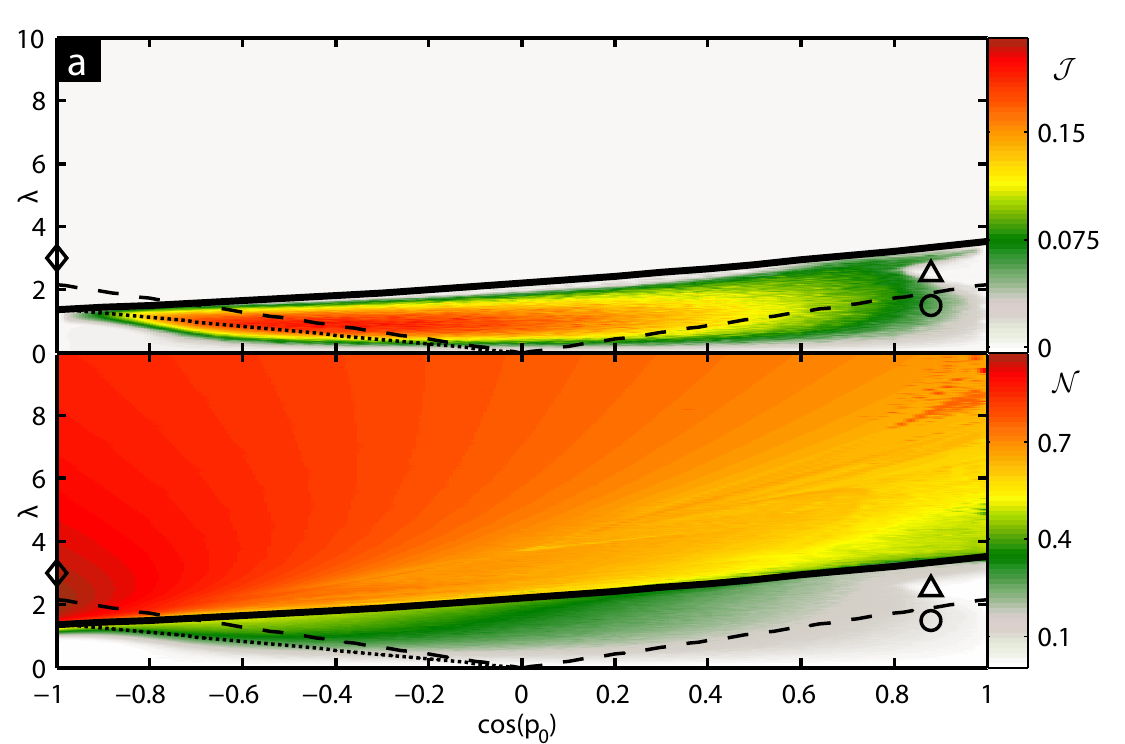}
      \includegraphics{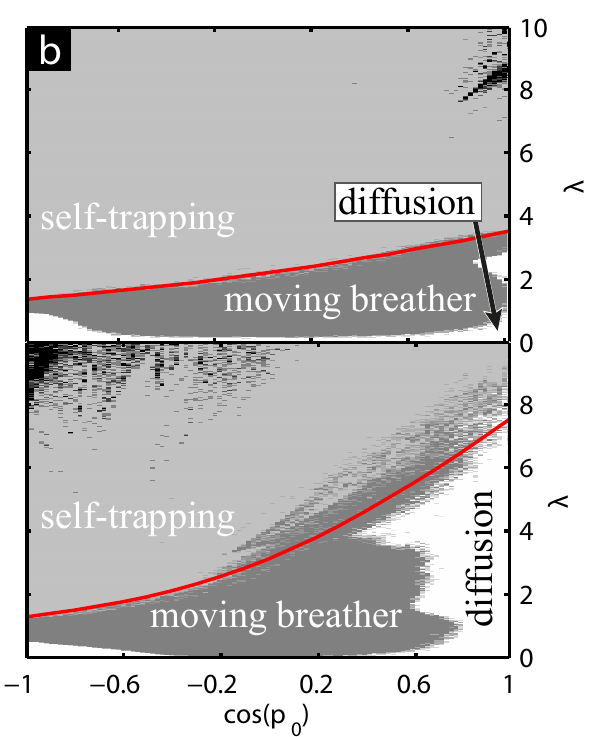}
      \caption{{\bf (a)} Construction of the dynamical phase diagram to identify diffusion, self-trapping and solitons based on two order parameters $\mathcal{J}$ (top) and $\mathcal{N}$ (bottom). We compare with our analytical estimate Eq.~\ref{lambda_critical} (thick black line) for the transition to self-trapping ($\alpha_0=1$). The variational approach \cite{Trombettoni:2001wl} (shown here as dashed and dotted lines) fails to predict the different regimes accurately:  above the dashed line all states were predicted to be self-trapped. In the region (for $\cos p<0$) between the dashed and the dotted line moving breathers should occur. The remainder would be the diffusive regime. In fact, we find that the dashed and dotted lines do not at all mark any of the transitions between the different dynamical regimes. Examples of the dynamics for three points in the phase diagram marked with the symbols
$\bigtriangleup$, $\bigcirc$, and $\Diamond$ are shown in Fig.~\ref{fig:dens00}.
{\bf (b)} Dynamical phase diagram of Gaussian initial conditions in the DNLS separating diffusion (white), self-trapping (light grey), moving breathers (dark grey) and higher order self-trapping (multi-breathers, black). The red curve Eq.~\ref{lambda_critical} represents our analytical estimate of the onset of self-trapping for
(top) $\alpha_0=1$ and $\alpha_0=4$ (bottom) and is in excellent agreement with the numerical data. Other parameters: $M=1001$, $\varepsilon=1.37$.
 }
      \label{fig:colorportrait_alpha1}
\end{figure*}

\emph{Model.}-- In the limit of large atom numbers per well, the dynamics of dilute Bose-Einstein condensates trapped in deep optical potentials are well described  by the mean-field Bose-Hubbard Hamiltonian \cite{Trombettoni:2001tb,PETHICK:2008tn,Buonsante:2008fe}
  \begin{align}
  \label{BHH}
    H=\sum\limits_{n=1}^{M} U|\psi_n|^4+\mu_n|\psi_n|^2-\frac{T}{2} \sum\limits_{n=1}^{M-1}  \psi_n^* \psi_{n+1}+c.c. ,
  \end{align}
 where $M$ is the lattice size, $|\psi_n(t)|^2$ is the norm (number of atoms at site $n$), $U$ denotes the onsite interaction (between two atoms at a single lattice site), $\mu_n$ is the onsite chemical potential and $T$ is the strength of the tunnel coupling between adjacent sites. The corresponding dynamical equation is the DNLS which read in its dimensionless form  
  \begin{align}
  \label{DNLS}
  i \frac{\partial \psi_n}{\partial t}= (\lambda |\psi_n|^2+\epsilon_n)\psi_n-\frac{1}{2} \left[ \psi_{n-1}+\psi_{n+1}\right]
  \end{align}
for $n=1...M$,  $\lambda=2U/T$ and $\epsilon_n=\mu_n/T$. In the numerics we present we use periodic boundary conditions ($\psi_{n+1}=\psi_1$), however, we checked that our results equally hold for closed boundary conditions. The DNLS describes a high dimensional chaotic dynamical system. Its dynamics, however, in general is far from being ergodic and e.g. shows localization in the form of stationary DB (localized excitations pinned to the lattice) and so called \emph{moving breathers} traversing the lattice~\cite{Flach:2008ud} (see Fig.~\ref{fig:dens00}).

The different characteristic types of dynamics are exemplified in Fig.~\ref{fig:dens00}: An initial condition that rapidly disperses, i.e.  shows diffusive behavior, can be seen in Fig.~\ref{fig:dens00}a. Moving breathers are strictly speaking not supported by the DNLS \cite{Oxtoby:2007gq}, they radiate norm and will eventually be pinned to the lattice or diffusively spread. However, as in Fig.~\ref{fig:dens00}b, these losses of norm can become infinitesimally small so that these solutions remain traveling for 
extremely long times and can therefore be regarded as actual moving breather solutions for practical use. Finally an example of a stationary DB emerging from a Gaussian initial condition is shown in Fig.~\ref{fig:dens00}c. 

Exact DB solutions were calculated analytically in \cite{Hennig:2010gy,Flach:2008ud} for a system of three sites (trimer). For larger lattices the DB solutions can be numerically calculated using the anti-continuous method 
\cite{Flach:2008ud,Aubry:1997wo,Marin:1996wv} or a Newton method \cite{Proville:1999vb}.
To prepare an experimental system in exact breather states is usually not feasible in practice.
In contrast, Gaussian initial conditions -- for which we derive a dynamical phase diagram --  can be well controlled in experiments on BECs with single-site addressability \cite{Gericke:2008tq, Albiez:2005wk,Morsch:2006em}. 

\emph{Variational approach.}-- In \cite{Trombettoni:2001wl,Trombettoni:2001tb} the dynamics of Gaussian wave packets defined by
 \begin{align}
 \label{ic}
    \psi_{n,0}\propto\exp\left[ \text{\small{$ -\frac{(n-\xi_0)^2}{\alpha_0}+i p_0 (n-\xi_0)+i\frac{\delta_0}{2}(n-\xi_0)^2$}} \right]
 \end{align}
was studied using a variational collective coordinate approach (a technique very successfully applied in a variety of fields 
\cite{Campbell:1983ta,Campbell:1980ua}). 
Here, $\xi_0$ and $\alpha_0$ are the center and the width of the Gaussian distribution and $p_0$ and $\delta_0$ their associated momenta. The variational approach leads to approximate equations of motion for the conjugate variables $(p, \xi)$ and $(\delta,\sqrt{\alpha})$ \cite{Trombettoni:2001wl,Trombettoni:2001tb}. It is important to note that the variational approach assumes that the excitation is well approximated by a Gaussian at \emph{all} times. The effective (approximate) Hamiltonian is~\cite{Trombettoni:2001tb}
 \begin{align}
 \label{AH}
 H=\frac{\lambda}{2\sqrt{\pi \alpha}}-\cos(p)e^{-\eta} 
 \end{align}
with $\eta=\frac{1}{2\alpha}+\frac{\alpha \delta^2}{8}$ depending on the initial values of the center $\xi_0$ and width parameter $\sqrt{\alpha_0}$  as well as $p_0$ and $\delta_0$ their conjugate momenta.

In ref.~\cite{Trombettoni:2001wl} a dynamical phase diagram was derived based on the variational approach Eq.~\ref{AH}.
These theoretical predictions however disagree with our numerical simulations of the actual DNLS dynamics, because the final dynamical state will in most cases be highly non-Gaussian. Hence, the variational approach breaks down, e.g.,
Fig.~\ref{fig:dens00}a shows diffusion, while the variational approach predicts a self-trapped state.
In Fig.~\ref{fig:dens00}b we find a moving breather while the variational approach yields diffusive behavior.
Moreover, the entire phase diagram Fig.~\ref{fig:colorportrait_alpha1}b(top) for the DNLS is completely different from the prediction of the variational approach (shown as dashed and dotted lines in Fig.~\ref{fig:colorportrait_alpha1}a).

\emph{Numerical construction of the phase diagram.}-- In order to construct a dynamical phase diagram based on the DNLS criteria to distinguish the different dynamical regimes are needed. This separation is non-trivial, as part of the atom cloud can, e.g., remain trapped in one region while another part diffuses in the remainder of the lattice. 
We can separate the regimes by defining a local norm and local average current
  \begin{align}
    N_{loc}(x)&=\sum\limits_{n=x-a}^{x+a} |\psi_n|^2 \notag \\
    j_{loc}(x)&=\frac{\hbar}{2m} \left| \sum\limits_{n=x-a}^{x+a} \textrm{Im}\left[\psi_{n+1}\psi_n^*-\psi_n^*\psi_{n-1}\right] \right|
  \end{align}
and two order parameters by
 \begin{align}
    \mathcal{N}=\langle \max_x(N_{loc}(x))\rangle \text{~~and~~}   \mathcal{J}=\langle j_{loc}(x_{max})\rangle,
    \label{eq:NJmax}
 \end{align}
where $x_{max}$ is the central site at which $N_{loc}(x)$ assumes its maximum. Both quantities were evaluated at time $\tau=10000$ and averaged over the last 10\% of the time.

In the diffusive regime both order parameters are small. Moving breathers are characterized by large $\mathcal{N}$ and large $\mathcal{J}$, while in the self-trapping regime one finds large $\mathcal{N}$ but nearly vanishing currents $\mathcal{J}$ due to the stationarity of the self-trapping solutions. Fig.~\ref{fig:colorportrait_alpha1} shows  $\mathcal{J}$ and $\mathcal{N}$ as a function of the initial phase difference $p_0$ and the nonlinearity $\lambda$ for $\alpha_0=1$ (see Supplemental Material for $\alpha_0=4$).
The examples of Fig.~\ref{fig:dens00} are marked in Fig.~\ref{fig:colorportrait_alpha1} by different symbols to demonstrate how the above criteria can be used to distinguish the dynamical regimes (a cut along $\cos (p_0)=0.88$ can be found in the Supplemental Material). 

We construct the dynamical phase diagram which assigns every initial condition to a specific dynamical regime by defining suitable thresholds for the order parameters $\mathcal{N}$ and $\mathcal{J}$. Figure~\ref{fig:colorportrait_alpha1}a(top) shows remarkably sharp transitions in the maximum local probability current. 
We use the thresholds $\mathcal{J}_{thrs}=0.002$ to separate moving breathers from self-trapping and diffusion and $\mathcal{N}_{
thrs}$ to delimit diffusion from self-trapping. By setting a threshold $p_T= 10^{-4}$ for the probability density function of $\mathcal{N}$ to be observed in a diffusive state (see Supplemental Material for details), we find $\mathcal{N}_{thrs}=0.028$ for $\alpha_0=1$ and $\mathcal{N}_{thrs}=0.032$ for $\alpha_0=4$.

With these thresholds we obtain the dynamical phase diagram shown in Fig.~\ref{fig:colorportrait_alpha1}b.
Our results are not sensitive to the exact value of the parameter $p_T$. For comparison we show a phase diagram obtained with $p_T = 10^{-5}$ in the Supplemental Material which differs only minimally from Fig.~\ref{fig:colorportrait_alpha1}b.
To identify higher order self-trapped excitations of more complex structure (see Fig.~\ref{fig:dens00}d) in our dynamical phase diagram, we use a third order parameter, the local energy 
$H_{loc}(x)=\sum_{n=x-a}^{x+a} \frac{\lambda}{2}|\psi_n|^4-    \sum_{n=x-a-1}^{x+a}\frac{1}{2} \left( \psi_n^* \psi_{n+1}+c.c.\right) \notag$ in the vicinity of $x_{max}$ measured relative to the fixed point energy $H_{cp}$ (which will be introduced below), i.e. $\mathcal{H}(\lambda)=H_{loc}(x_{\mathrm{max}})/H_{cp}(\lambda,1)$.
For higher order self-trapping we found $\mathcal{H}\ll 1$ in contrast to $\mathcal{H}\approx 1$ for discrete breathers centered around a single site. We chose $\mathcal{H}_{thrs}=0.8$ and marked this regime in black in Fig.~\ref{fig:colorportrait_alpha1}b. 

\emph{Theory.}-- In the following, we derive an analytical estimate for the self-trapping transition separating the different phases in the dynamical phase diagram.
The exact DB solution forming a single peak centered at a single lattice site (denoted  \textit{cp} for \textit{central peak}) is a local maximum in the energy and can be numerically constructed using the methods described in Refs.~\cite{Proville:1999vb,Flach:2008ud,Aubry:1997wo,Marin:1996wv}. The cp-solutions correspond to elliptic fixed points in phase space \cite{Rumpf:2004en}
which are separated from moving and chaotic solutions by saddle points in the energy landscape. These saddle points correspond to another kind of stationary solutions centered between two lattice sites called \textit{central bond} (cb) states. They are unstable \cite{Rumpf:2004en} and have lower energy than cp-solutions. Due to continuity the cb-solutions are considered intermediate states for an excitation to hop between lattice sites. 

A necessary condition for self-trapping is $H_{G} \ge H_{cb}$, where $H_{G}$ is the energy of the initial Gaussian and $H_{cb}$ is the energy of the cb-solution.
To derive an expression for $H_{cb}$, note that the energy of an arbitrary state of norm $n$ compared to the state of the same shape but of unit norm scales as $E(\lambda,n)=n\cdot E(\lambda \cdot n,1)$.
Thus $H_{cb}$ reads
\begin{equation}
H_{cb}(\lambda,n)=n\cdot H_{cb}(\lambda \cdot n,1)=\frac{\Lambda}{\lambda}\cdot H_{cb}(\Lambda,1)
\label{Hcb}
\end{equation}
with $\Lambda=\lambda \cdot n$.
We approximate the energy of the initial Gaussian by $H_{G}=\frac{\lambda}{2\sqrt{\pi \alpha_0}}-\cos(p_0) e^{-\eta_0} $ with $\eta_0=\frac{1}{2 \alpha_0}$ (see Eq.~\ref{AH}), which we found agrees very well with the energy $H$ of the DNLS Hamiltonian (Eq.~(\ref{BHH})), although the dynamics strongly differs. 

We make the following Ansatz to estimate the critical nonlinearity $\lambda_c$ at the transition to self trapping
 \begin{align}
   H_{cb}(\lambda_c,n_c)&=\frac{\Lambda_c}{\lambda_c}\cdot H_{cb}(\Lambda_c,1)=\frac{\varepsilon}{\lambda_c}\stackrel{!}{=}H_{G}(\lambda_c) 
 \label{Hcritical}
 \end{align}
where $\varepsilon=\Lambda_c \cdot H_{cb}(\Lambda_c,1)$ and $\Lambda_c = \lambda_c n_c$. 
The critical nonlinearity at the transition to self-trapping reads
\begin{eqnarray}
\label{lambda_critical}
\lambda_c(p_0) 
& = & e^{-\eta} \sqrt{ \pi \alpha_0} \cos p_0 \nonumber \\
& &
+ \sqrt{e^{-2\eta} \pi \alpha_0 \cos^2 p_0 +2 \varepsilon \sqrt{\pi \alpha_0} }
 \end{eqnarray}  

Our analytical estimate of the self-trapping transition Eq.~(\ref{lambda_critical}) is shown in Fig.~\ref{fig:colorportrait_alpha1}a (thick black line) and Fig.~\ref{fig:colorportrait_alpha1}b (thick red line) 
and is in excellent agreement with the simulations of the DNLS dynamics.
By fixing $p_0=\pi$ we can determine the constant $\varepsilon$ numerically. We find $\lambda_{c}(p_0) \approx 1.35$ and $\varepsilon_\text{num}=1.375$. We chose $p_0 = \pi$ for the estimate since the phase difference between neighboring sites for an exact solution of a discrete breather is $\pi$. Below, we will present an independent way to obtain $\varepsilon$ by calculating $\Lambda_c$ directly.

How can we interpret the rescaled nonlinearity $\Lambda_c$ in Eq.~\ref{Hcritical}?
The ability of localized excitations to move inside a lattice can be analyzed using the concept of the Peierls-Nabarro (PN) energy barrier which we shortly review: A crucial condition for a localized excitation to move across the lattice is that the initial energy is lower than the energy of a cb-solution such that this intermediate state cannot act as a barrier. Self-trapping on the other hand requires the initial energy to be between the maximum energy state cp and cb which can thus act as a barrier in phase space and inhibits migration. This energy gap $H_{PN}$ between cp and cb is referred to as the PN barrier \cite{Kivshar:1993vf, Hennig:2010gy, Rumpf:2004en}. 

\begin{figure}[h!]
    \centering
    \includegraphics[width=0.8\columnwidth]{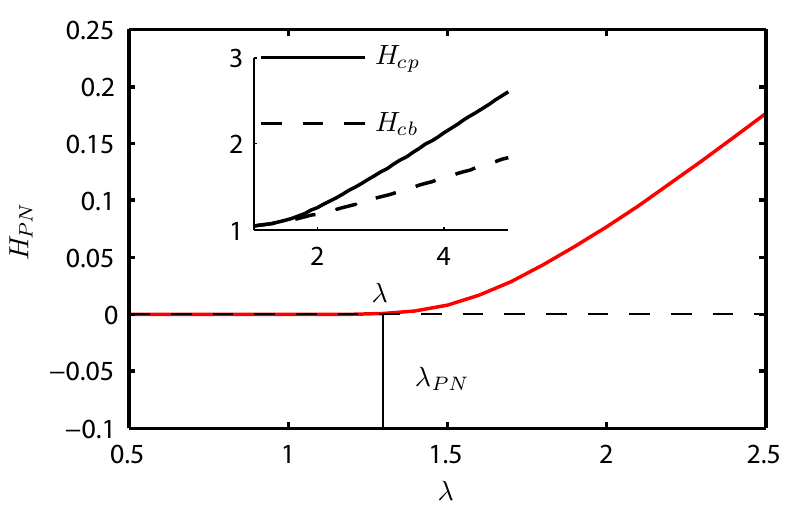}
    \caption{The onset of a non-zero PN barrier $H_{PN}$ is found at nonlinearity $\lambda_{PN}\approx1.3$ (for $n=1$). THe PN barrier $H_{PN}$ is calculated as the difference between the energies of a cp breather (located around a single site) and a cb breather (located between two lattice sites).}
    \label{fig:pn_barrier}
 \end{figure}

We report the PN barrier in Fig.~\ref{fig:pn_barrier}. Self-trapping requires a non-zero PN barrier, which we find for nonlinearities larger than $\lambda_{PN}\approx1.3$ in Fig.~\ref{fig:pn_barrier}. Since the transition to self-trapping is reflected by the onset of a non-zero PN barrier, we identify $\Lambda_C \equiv \lambda_{PN}$. With $H_{cb}(\lambda_{PN},1)\approx1.0723$ we find 
$\varepsilon=1.37$ in excellent agreement with the numerical value $\varepsilon_\text{num}$.
Note that only a fraction of the norm of the initial Gaussian will actually be trapped when a breather is created (and this norm is smaller the more the phase differences $p_0$ of the Gaussian deviates from the phase difference of the breather fixed point, i.e. $p=\pi$).
Our theoretical analysis focuses on the self-trapped fraction of norm and thus the existence of a non-zero PN barrier corresponding to the respective fraction of norm $n$. This of course leaves the background unattended yet proves to be a valid approximation.

In conclusion, we calculated a dynamical phase diagram for the different dynamical regimes (diffusion, moving breathers and self-trapping) of Gaussian initial conditions in periodic optical potentials. By defining two order parameters -- the maximum local atomic density and the maximum local current -- we numerically construct the dynamical phase diagram. We derived an explicit expression (Eq.~\ref{lambda_critical}) for the nonlinear interaction $\lambda_c$ that separates the dynamical regimes, in very good agreement with the numerical results.
Gaussian initial conditions can be well controlled in experiments on BECs with single-site adressability \cite{Gericke:2008tq, Albiez:2005wk,Morsch:2006em} where our predictions can be readily tested experimentally. 
We hope our study further stimulates experiments and theoretical work on the transition between self-trapping, diffusion and solitons in the quantum regime, e.g., based on the Bose-Hubbard Hamiltonian.

\begin{acknowledgments}
HH acknowledges financial support by the German Research Foundation
(DFG), grant no.~HE 6312/1-1. We thank J\'{e}r\^{o}me Dorignac and David K.~Campbell for useful discussions. 
\end{acknowledgments}

%\bibliographystyle{apsrev}
%\bibliography{papers_hh}

\begin{thebibliography}{50}
\expandafter\ifx\csname natexlab\endcsname\relax\def\natexlab#1{#1}\fi
\expandafter\ifx\csname bibnamefont\endcsname\relax
  \def\bibnamefont#1{#1}\fi
\expandafter\ifx\csname bibfnamefont\endcsname\relax
  \def\bibfnamefont#1{#1}\fi
\expandafter\ifx\csname citenamefont\endcsname\relax
  \def\citenamefont#1{#1}\fi
\expandafter\ifx\csname url\endcsname\relax
  \def\url#1{\texttt{#1}}\fi
\expandafter\ifx\csname urlprefix\endcsname\relax\def\urlprefix{URL }\fi
\providecommand{\bibinfo}[2]{#2}
\providecommand{\eprint}[2][]{\url{#2}}

\bibitem[{\citenamefont{Bloch}(2005)}]{Bloch:2005uv}
\bibinfo{author}{\bibfnamefont{I.}~\bibnamefont{Bloch}}, \bibinfo{journal}{Nat.
  Phys.} \textbf{\bibinfo{volume}{1}}, \bibinfo{pages}{23}
  (\bibinfo{year}{2005}).

\bibitem[{\citenamefont{Bakr et~al.}(2009)\citenamefont{Bakr, Gillen, Peng,
  F{\"o}lling, and Greiner}}]{Bakr:2009bx}
\bibinfo{author}{\bibfnamefont{W.~S.} \bibnamefont{Bakr}},
  \bibinfo{author}{\bibfnamefont{J.~I.} \bibnamefont{Gillen}},
  \bibinfo{author}{\bibfnamefont{A.}~\bibnamefont{Peng}},
  \bibinfo{author}{\bibfnamefont{S.}~\bibnamefont{F{\"o}lling}},
  \bibnamefont{and} \bibinfo{author}{\bibfnamefont{M.}~\bibnamefont{Greiner}},
  \bibinfo{journal}{Nature} \textbf{\bibinfo{volume}{461}}, \bibinfo{pages}{74}
  (\bibinfo{year}{2009}).

\bibitem[{\citenamefont{Bakr et~al.}(2010)\citenamefont{Bakr, Peng, Tai, Ma,
  Simon, Gillen, F{\"o}lling, Pollet, and Greiner}}]{Bakr:2010gd}
\bibinfo{author}{\bibfnamefont{W.~S.} \bibnamefont{Bakr}},
  \bibinfo{author}{\bibfnamefont{A.}~\bibnamefont{Peng}},
  \bibinfo{author}{\bibfnamefont{M.~E.} \bibnamefont{Tai}},
  \bibinfo{author}{\bibfnamefont{R.}~\bibnamefont{Ma}},
  \bibinfo{author}{\bibfnamefont{J.}~\bibnamefont{Simon}},
  \bibinfo{author}{\bibfnamefont{J.~I.} \bibnamefont{Gillen}},
  \bibinfo{author}{\bibfnamefont{S.}~\bibnamefont{F{\"o}lling}},
  \bibinfo{author}{\bibfnamefont{L.}~\bibnamefont{Pollet}}, \bibnamefont{and}
  \bibinfo{author}{\bibfnamefont{M.}~\bibnamefont{Greiner}},
  \bibinfo{journal}{Science} \textbf{\bibinfo{volume}{329}},
  \bibinfo{pages}{547} (\bibinfo{year}{2010}).

\bibitem[{\citenamefont{Simon et~al.}(2011)\citenamefont{Simon, Bakr, Ma, Tai,
  Preiss, and Greiner}}]{Simon:2011ep}
\bibinfo{author}{\bibfnamefont{J.}~\bibnamefont{Simon}},
  \bibinfo{author}{\bibfnamefont{W.~S.} \bibnamefont{Bakr}},
  \bibinfo{author}{\bibfnamefont{R.}~\bibnamefont{Ma}},
  \bibinfo{author}{\bibfnamefont{M.~E.} \bibnamefont{Tai}},
  \bibinfo{author}{\bibfnamefont{P.~M.} \bibnamefont{Preiss}},
  \bibnamefont{and} \bibinfo{author}{\bibfnamefont{M.}~\bibnamefont{Greiner}},
  \bibinfo{journal}{Nature} \textbf{\bibinfo{volume}{472}},
  \bibinfo{pages}{307} (\bibinfo{year}{2011}).

\bibitem[{\citenamefont{Morsch and Oberthaler}(2006)}]{Morsch:2006em}
\bibinfo{author}{\bibfnamefont{O.}~\bibnamefont{Morsch}} \bibnamefont{and}
  \bibinfo{author}{\bibfnamefont{M.}~\bibnamefont{Oberthaler}},
  \bibinfo{journal}{Rev. Mod. Phys.} \textbf{\bibinfo{volume}{78}},
  \bibinfo{pages}{179} (\bibinfo{year}{2006}).

\bibitem[{\citenamefont{Pethick and Smith}(2008)}]{PETHICK:2008tn}
\bibinfo{author}{\bibfnamefont{C.~J.} \bibnamefont{Pethick}} \bibnamefont{and}
  \bibinfo{author}{\bibfnamefont{H.}~\bibnamefont{Smith}},
  \emph{\bibinfo{title}{{Bose-Einstein Condensation in Dilute Gases}}}
  (\bibinfo{publisher}{Cambridge University Press},
  \bibinfo{address}{Cambridge, UK}, \bibinfo{year}{2008}).

\bibitem[{\citenamefont{Flach and Gorbach}(2008)}]{Flach:2008ud}
\bibinfo{author}{\bibfnamefont{S.}~\bibnamefont{Flach}} \bibnamefont{and}
  \bibinfo{author}{\bibfnamefont{A.~V.} \bibnamefont{Gorbach}},
  \bibinfo{journal}{Phys. Rep.} \textbf{\bibinfo{volume}{467}},
  \bibinfo{pages}{1} (\bibinfo{year}{2008}).

\bibitem[{\citenamefont{Campbell et~al.}(2004)\citenamefont{Campbell, Flach,
  and Kivshar}}]{Campbell:2004vz}
\bibinfo{author}{\bibfnamefont{D.~K.} \bibnamefont{Campbell}},
  \bibinfo{author}{\bibfnamefont{S.}~\bibnamefont{Flach}}, \bibnamefont{and}
  \bibinfo{author}{\bibfnamefont{Y.~S.} \bibnamefont{Kivshar}},
  \bibinfo{journal}{Phys. Today} \textbf{\bibinfo{volume}{57}},
  \bibinfo{pages}{43} (\bibinfo{year}{2004}).

\bibitem[{\citenamefont{Hennig and Fleischmann}(2013)}]{Hennig:2013ja}
\bibinfo{author}{\bibfnamefont{H.}~\bibnamefont{Hennig}} \bibnamefont{and}
  \bibinfo{author}{\bibfnamefont{R.}~\bibnamefont{Fleischmann}},
  \bibinfo{journal}{Phys. Rev. A} \textbf{\bibinfo{volume}{87}},
  \bibinfo{pages}{033605} (\bibinfo{year}{2013}).

\bibitem[{\citenamefont{Eisenberg et~al.}(1998)\citenamefont{Eisenberg,
  Silberberg, Morandotti, Boyd, and Aitchison}}]{Eisenberg:1998wp}
\bibinfo{author}{\bibfnamefont{H.~S.} \bibnamefont{Eisenberg}},
  \bibinfo{author}{\bibfnamefont{Y.}~\bibnamefont{Silberberg}},
  \bibinfo{author}{\bibfnamefont{R.}~\bibnamefont{Morandotti}},
  \bibinfo{author}{\bibfnamefont{A.~R.} \bibnamefont{Boyd}}, \bibnamefont{and}
  \bibinfo{author}{\bibfnamefont{J.~S.} \bibnamefont{Aitchison}},
  \bibinfo{journal}{Phys. Rev. Lett.} \textbf{\bibinfo{volume}{81}},
  \bibinfo{pages}{3383} (\bibinfo{year}{1998}).

\bibitem[{\citenamefont{Morandotti et~al.}(1999)\citenamefont{Morandotti,
  Peschel, Aitchison, Eisenberg, and Silberberg}}]{Morandotti:1999tt}
\bibinfo{author}{\bibfnamefont{R.}~\bibnamefont{Morandotti}},
  \bibinfo{author}{\bibfnamefont{U.}~\bibnamefont{Peschel}},
  \bibinfo{author}{\bibfnamefont{J.~S.} \bibnamefont{Aitchison}},
  \bibinfo{author}{\bibfnamefont{H.~S.} \bibnamefont{Eisenberg}},
  \bibnamefont{and}
  \bibinfo{author}{\bibfnamefont{Y.}~\bibnamefont{Silberberg}},
  \bibinfo{journal}{Phys. Rev. Lett.} \textbf{\bibinfo{volume}{83}},
  \bibinfo{pages}{2726} (\bibinfo{year}{1999}).

\bibitem[{\citenamefont{Tr{\'\i}as et~al.}(2000)\citenamefont{Tr{\'\i}as, Mazo,
  and Orlando}}]{Trias:2000uw}
\bibinfo{author}{\bibfnamefont{E.}~\bibnamefont{Tr{\'\i}as}},
  \bibinfo{author}{\bibfnamefont{J.~J.} \bibnamefont{Mazo}}, \bibnamefont{and}
  \bibinfo{author}{\bibfnamefont{T.~P.} \bibnamefont{Orlando}},
  \bibinfo{journal}{Phys. Rev. Lett.} \textbf{\bibinfo{volume}{84}},
  \bibinfo{pages}{741} (\bibinfo{year}{2000}).

\bibitem[{\citenamefont{Ustinov}(2003)}]{Ustinov:2003uc}
\bibinfo{author}{\bibfnamefont{A.~V.} \bibnamefont{Ustinov}},
  \bibinfo{journal}{Chaos} \textbf{\bibinfo{volume}{13}}, \bibinfo{pages}{716}
  (\bibinfo{year}{2003}).

\bibitem[{\citenamefont{Schwarz et~al.}(1999)\citenamefont{Schwarz, English,
  and Sievers}}]{Schwarz:1999vb}
\bibinfo{author}{\bibfnamefont{U.~T.} \bibnamefont{Schwarz}},
  \bibinfo{author}{\bibfnamefont{L.~Q.} \bibnamefont{English}},
  \bibnamefont{and} \bibinfo{author}{\bibfnamefont{A.~J.}
  \bibnamefont{Sievers}}, \bibinfo{journal}{Phys. Rev. Lett.}
  \textbf{\bibinfo{volume}{83}}, \bibinfo{pages}{223} (\bibinfo{year}{1999}).

\bibitem[{\citenamefont{Sato and Sievers}(2004)}]{Sato:2004vw}
\bibinfo{author}{\bibfnamefont{M.}~\bibnamefont{Sato}} \bibnamefont{and}
  \bibinfo{author}{\bibfnamefont{A.~J.} \bibnamefont{Sievers}},
  \bibinfo{journal}{Nature} \textbf{\bibinfo{volume}{432}},
  \bibinfo{pages}{486} (\bibinfo{year}{2004}).

\bibitem[{\citenamefont{Eiermann et~al.}(2004)\citenamefont{Eiermann, Anker,
  Albiez, Taglieber, Treutlein, Marzlin, and Oberthaler}}]{Eiermann:2004vt}
\bibinfo{author}{\bibfnamefont{B.}~\bibnamefont{Eiermann}},
  \bibinfo{author}{\bibfnamefont{T.}~\bibnamefont{Anker}},
  \bibinfo{author}{\bibfnamefont{M.}~\bibnamefont{Albiez}},
  \bibinfo{author}{\bibfnamefont{M.}~\bibnamefont{Taglieber}},
  \bibinfo{author}{\bibfnamefont{P.}~\bibnamefont{Treutlein}},
  \bibinfo{author}{\bibfnamefont{K.-P.} \bibnamefont{Marzlin}},
  \bibnamefont{and} \bibinfo{author}{\bibfnamefont{M.~K.}
  \bibnamefont{Oberthaler}}, \bibinfo{journal}{Phys. Rev. Lett.}
  \textbf{\bibinfo{volume}{92}}, \bibinfo{pages}{230401}
  (\bibinfo{year}{2004}).

\bibitem[{\citenamefont{Hennig et~al.}(2012)\citenamefont{Hennig, Witthaut, and
  Campbell}}]{Hennig:2012gn}
\bibinfo{author}{\bibfnamefont{H.}~\bibnamefont{Hennig}},
  \bibinfo{author}{\bibfnamefont{D.}~\bibnamefont{Witthaut}}, \bibnamefont{and}
  \bibinfo{author}{\bibfnamefont{D.}~\bibnamefont{Campbell}},
  \bibinfo{journal}{Phys. Rev. A} \textbf{\bibinfo{volume}{86}},
  \bibinfo{pages}{051604} (\bibinfo{year}{2012}).

\bibitem[{\citenamefont{Sakmann et~al.}(2009)\citenamefont{Sakmann, Streltsov,
  Alon, and Cederbaum}}]{Sakmann:2009wy}
\bibinfo{author}{\bibfnamefont{K.}~\bibnamefont{Sakmann}},
  \bibinfo{author}{\bibfnamefont{A.~I.} \bibnamefont{Streltsov}},
  \bibinfo{author}{\bibfnamefont{O.}~\bibnamefont{Alon}}, \bibnamefont{and}
  \bibinfo{author}{\bibfnamefont{L.~S.} \bibnamefont{Cederbaum}},
  \bibinfo{journal}{Phys. Rev. Lett.} \textbf{\bibinfo{volume}{103}},
  \bibinfo{pages}{220601} (\bibinfo{year}{2009}).

\bibitem[{\citenamefont{Trujillo-Martinez
  et~al.}(2009)\citenamefont{Trujillo-Martinez, Posazhennikova, and
  Kroha}}]{TrujilloMartinez:2009gh}
\bibinfo{author}{\bibfnamefont{M.}~\bibnamefont{Trujillo-Martinez}},
  \bibinfo{author}{\bibfnamefont{A.}~\bibnamefont{Posazhennikova}},
  \bibnamefont{and} \bibinfo{author}{\bibfnamefont{J.}~\bibnamefont{Kroha}},
  \bibinfo{journal}{Phys. Rev. Lett.} \textbf{\bibinfo{volume}{103}},
  \bibinfo{pages}{105302} (\bibinfo{year}{2009}).

\bibitem[{\citenamefont{Trombettoni and
  Smerzi}(2001{\natexlab{a}})}]{Trombettoni:2001wl}
\bibinfo{author}{\bibfnamefont{A.}~\bibnamefont{Trombettoni}} \bibnamefont{and}
  \bibinfo{author}{\bibfnamefont{A.}~\bibnamefont{Smerzi}},
  \bibinfo{journal}{Phys. Rev. Lett.} \textbf{\bibinfo{volume}{86}},
  \bibinfo{pages}{2353} (\bibinfo{year}{2001}{\natexlab{a}}).

\bibitem[{\citenamefont{Christodoulides
  et~al.}(2003)\citenamefont{Christodoulides, Lederer, and
  Silberberg}}]{Christodoulides:2003vu}
\bibinfo{author}{\bibfnamefont{D.~N.} \bibnamefont{Christodoulides}},
  \bibinfo{author}{\bibfnamefont{F.}~\bibnamefont{Lederer}}, \bibnamefont{and}
  \bibinfo{author}{\bibfnamefont{Y.}~\bibnamefont{Silberberg}},
  \bibinfo{journal}{Nature} \textbf{\bibinfo{volume}{424}},
  \bibinfo{pages}{817} (\bibinfo{year}{2003}).

\bibitem[{\citenamefont{Fleischer
  et~al.}(2003{\natexlab{a}})\citenamefont{Fleischer, Segev, Efremidis, and
  Christodoulides}}]{Fleischer:2003gc}
\bibinfo{author}{\bibfnamefont{J.~W.} \bibnamefont{Fleischer}},
  \bibinfo{author}{\bibfnamefont{M.}~\bibnamefont{Segev}},
  \bibinfo{author}{\bibfnamefont{N.~K.} \bibnamefont{Efremidis}},
  \bibnamefont{and} \bibinfo{author}{\bibfnamefont{D.~N.}
  \bibnamefont{Christodoulides}}, \bibinfo{journal}{Nature}
  \textbf{\bibinfo{volume}{422}}, \bibinfo{pages}{147}
  (\bibinfo{year}{2003}{\natexlab{a}}).

\bibitem[{\citenamefont{Albiez et~al.}(2005)\citenamefont{Albiez, Gati,
  F{\"o}lling, Hunsmann, Cristiani, and Oberthaler}}]{Albiez:2005wk}
\bibinfo{author}{\bibfnamefont{M.}~\bibnamefont{Albiez}},
  \bibinfo{author}{\bibfnamefont{R.}~\bibnamefont{Gati}},
  \bibinfo{author}{\bibfnamefont{J.}~\bibnamefont{F{\"o}lling}},
  \bibinfo{author}{\bibfnamefont{S.}~\bibnamefont{Hunsmann}},
  \bibinfo{author}{\bibfnamefont{M.}~\bibnamefont{Cristiani}},
  \bibnamefont{and}
  \bibinfo{author}{\bibfnamefont{M.}~\bibnamefont{Oberthaler}},
  \bibinfo{journal}{Phys. Rev. Lett.} \textbf{\bibinfo{volume}{95}},
  \bibinfo{pages}{010402} (\bibinfo{year}{2005}).

\bibitem[{\citenamefont{Cataliotti et~al.}(2001)\citenamefont{Cataliotti,
  Burger, Fort, Maddaloni, Minardi, Trombettoni, Smerzi, and
  Inguscio}}]{Cataliotti:2001tx}
\bibinfo{author}{\bibfnamefont{F.~S.} \bibnamefont{Cataliotti}},
  \bibinfo{author}{\bibfnamefont{S.}~\bibnamefont{Burger}},
  \bibinfo{author}{\bibfnamefont{C.}~\bibnamefont{Fort}},
  \bibinfo{author}{\bibfnamefont{P.}~\bibnamefont{Maddaloni}},
  \bibinfo{author}{\bibfnamefont{F.}~\bibnamefont{Minardi}},
  \bibinfo{author}{\bibfnamefont{A.}~\bibnamefont{Trombettoni}},
  \bibinfo{author}{\bibfnamefont{A.}~\bibnamefont{Smerzi}}, \bibnamefont{and}
  \bibinfo{author}{\bibfnamefont{M.}~\bibnamefont{Inguscio}},
  \bibinfo{journal}{Science} \textbf{\bibinfo{volume}{293}},
  \bibinfo{pages}{843} (\bibinfo{year}{2001}).

\bibitem[{\citenamefont{Fleischer
  et~al.}(2003{\natexlab{b}})\citenamefont{Fleischer, Fleischer, Carmon,
  Carmon, Segev, Segev, Efremidis, Efremidis, Christodoulides, and
  Christodoulides}}]{Fleischer:2003cr}
\bibinfo{author}{\bibfnamefont{J.}~\bibnamefont{Fleischer}},
  \bibinfo{author}{\bibfnamefont{J.}~\bibnamefont{Fleischer}},
  \bibinfo{author}{\bibfnamefont{T.}~\bibnamefont{Carmon}},
  \bibinfo{author}{\bibfnamefont{T.}~\bibnamefont{Carmon}},
  \bibinfo{author}{\bibfnamefont{M.}~\bibnamefont{Segev}},
  \bibinfo{author}{\bibfnamefont{M.}~\bibnamefont{Segev}},
  \bibinfo{author}{\bibfnamefont{N.}~\bibnamefont{Efremidis}},
  \bibinfo{author}{\bibfnamefont{N.}~\bibnamefont{Efremidis}},
  \bibinfo{author}{\bibfnamefont{D.}~\bibnamefont{Christodoulides}},
  \bibnamefont{and}
  \bibinfo{author}{\bibfnamefont{D.}~\bibnamefont{Christodoulides}},
  \bibinfo{journal}{Phys. Rev. Lett.} \textbf{\bibinfo{volume}{90}},
  \bibinfo{pages}{023902} (\bibinfo{year}{2003}{\natexlab{b}}).

\bibitem[{\citenamefont{Morsch et~al.}(2001)\citenamefont{Morsch, M{\"u}ller,
  Cristiani, Ciampini, and Arimondo}}]{Morsch:2001hj}
\bibinfo{author}{\bibfnamefont{O.}~\bibnamefont{Morsch}},
  \bibinfo{author}{\bibfnamefont{J.}~\bibnamefont{M{\"u}ller}},
  \bibinfo{author}{\bibfnamefont{M.}~\bibnamefont{Cristiani}},
  \bibinfo{author}{\bibfnamefont{D.}~\bibnamefont{Ciampini}}, \bibnamefont{and}
  \bibinfo{author}{\bibfnamefont{E.}~\bibnamefont{Arimondo}},
  \bibinfo{journal}{Phys. Rev. Lett.} \textbf{\bibinfo{volume}{87}},
  \bibinfo{pages}{140402} (\bibinfo{year}{2001}).

\bibitem[{\citenamefont{Burger et~al.}(1999)\citenamefont{Burger, Bongs,
  Dettmer, Ertmer, Sengstock, Sanpera, Shlyapnikov, and
  Lewenstein}}]{Burger:1999vs}
\bibinfo{author}{\bibfnamefont{S.}~\bibnamefont{Burger}},
  \bibinfo{author}{\bibfnamefont{K.}~\bibnamefont{Bongs}},
  \bibinfo{author}{\bibfnamefont{S.}~\bibnamefont{Dettmer}},
  \bibinfo{author}{\bibfnamefont{W.}~\bibnamefont{Ertmer}},
  \bibinfo{author}{\bibfnamefont{K.}~\bibnamefont{Sengstock}},
  \bibinfo{author}{\bibfnamefont{A.}~\bibnamefont{Sanpera}},
  \bibinfo{author}{\bibfnamefont{G.~V.} \bibnamefont{Shlyapnikov}},
  \bibnamefont{and}
  \bibinfo{author}{\bibfnamefont{M.}~\bibnamefont{Lewenstein}},
  \bibinfo{journal}{Phys. Rev. Lett.} \textbf{\bibinfo{volume}{83}},
  \bibinfo{pages}{5198} (\bibinfo{year}{1999}).

\bibitem[{\citenamefont{Smerzi et~al.}(2002)\citenamefont{Smerzi, Trombettoni,
  Kevrekidis, and Bishop}}]{Smerzi:2002tj}
\bibinfo{author}{\bibfnamefont{A.}~\bibnamefont{Smerzi}},
  \bibinfo{author}{\bibfnamefont{A.}~\bibnamefont{Trombettoni}},
  \bibinfo{author}{\bibfnamefont{P.}~\bibnamefont{Kevrekidis}},
  \bibnamefont{and} \bibinfo{author}{\bibfnamefont{A.}~\bibnamefont{Bishop}},
  \bibinfo{journal}{Phys. Rev. Lett.} \textbf{\bibinfo{volume}{89}},
  \bibinfo{pages}{170402} (\bibinfo{year}{2002}).

\bibitem[{\citenamefont{Anker et~al.}(2005)\citenamefont{Anker, Albiez, Gati,
  Hunsmann, Eiermann, Trombettoni, and Oberthaler}}]{Anker:2005vf}
\bibinfo{author}{\bibfnamefont{T.}~\bibnamefont{Anker}},
  \bibinfo{author}{\bibfnamefont{M.}~\bibnamefont{Albiez}},
  \bibinfo{author}{\bibfnamefont{R.}~\bibnamefont{Gati}},
  \bibinfo{author}{\bibfnamefont{S.}~\bibnamefont{Hunsmann}},
  \bibinfo{author}{\bibfnamefont{B.}~\bibnamefont{Eiermann}},
  \bibinfo{author}{\bibfnamefont{A.}~\bibnamefont{Trombettoni}},
  \bibnamefont{and}
  \bibinfo{author}{\bibfnamefont{M.}~\bibnamefont{Oberthaler}},
  \bibinfo{journal}{Phys. Rev. Lett.} \textbf{\bibinfo{volume}{94}},
  \bibinfo{pages}{020403} (\bibinfo{year}{2005}).

\bibitem[{\citenamefont{Iwanow et~al.}(2004)\citenamefont{Iwanow, Schiek,
  Stegeman, Pertsch, Lederer, Min, and Sohler}}]{Iwanow:2004df}
\bibinfo{author}{\bibfnamefont{R.}~\bibnamefont{Iwanow}},
  \bibinfo{author}{\bibfnamefont{R.}~\bibnamefont{Schiek}},
  \bibinfo{author}{\bibfnamefont{G.}~\bibnamefont{Stegeman}},
  \bibinfo{author}{\bibfnamefont{T.}~\bibnamefont{Pertsch}},
  \bibinfo{author}{\bibfnamefont{F.}~\bibnamefont{Lederer}},
  \bibinfo{author}{\bibfnamefont{Y.}~\bibnamefont{Min}}, \bibnamefont{and}
  \bibinfo{author}{\bibfnamefont{W.}~\bibnamefont{Sohler}},
  \bibinfo{journal}{Phys. Rev. Lett.} \textbf{\bibinfo{volume}{93}},
  \bibinfo{pages}{113902} (\bibinfo{year}{2004}).

\bibitem[{\citenamefont{Eiermann et~al.}(2003)\citenamefont{Eiermann,
  Treutlein, Anker, Albiez, Taglieber, Marzlin, and
  Oberthaler}}]{Eiermann:2003ja}
\bibinfo{author}{\bibfnamefont{B.}~\bibnamefont{Eiermann}},
  \bibinfo{author}{\bibfnamefont{P.}~\bibnamefont{Treutlein}},
  \bibinfo{author}{\bibfnamefont{T.}~\bibnamefont{Anker}},
  \bibinfo{author}{\bibfnamefont{M.}~\bibnamefont{Albiez}},
  \bibinfo{author}{\bibfnamefont{M.}~\bibnamefont{Taglieber}},
  \bibinfo{author}{\bibfnamefont{K.-P.} \bibnamefont{Marzlin}},
  \bibnamefont{and}
  \bibinfo{author}{\bibfnamefont{M.}~\bibnamefont{Oberthaler}},
  \bibinfo{journal}{Phys. Rev. Lett.} \textbf{\bibinfo{volume}{91}},
  \bibinfo{pages}{060402} (\bibinfo{year}{2003}).

\bibitem[{\citenamefont{Reinhard et~al.}(2013)\citenamefont{Reinhard, Riou,
  Zundel, Weiss, Li, Rey, and Hipolito}}]{Reinhard:2013ft}
\bibinfo{author}{\bibfnamefont{A.}~\bibnamefont{Reinhard}},
  \bibinfo{author}{\bibfnamefont{J.-F.} \bibnamefont{Riou}},
  \bibinfo{author}{\bibfnamefont{L.}~\bibnamefont{Zundel}},
  \bibinfo{author}{\bibfnamefont{D.}~\bibnamefont{Weiss}},
  \bibinfo{author}{\bibfnamefont{S.}~\bibnamefont{Li}},
  \bibinfo{author}{\bibfnamefont{A.}~\bibnamefont{Rey}}, \bibnamefont{and}
  \bibinfo{author}{\bibfnamefont{R.}~\bibnamefont{Hipolito}},
  \bibinfo{journal}{Phys. Rev. Lett.} \textbf{\bibinfo{volume}{110}},
  \bibinfo{pages}{033001} (\bibinfo{year}{2013}).

\bibitem[{\citenamefont{Bersch et~al.}(2012)\citenamefont{Bersch, Onishchukov,
  and Peschel}}]{Bersch:2012ih}
\bibinfo{author}{\bibfnamefont{C.}~\bibnamefont{Bersch}},
  \bibinfo{author}{\bibfnamefont{G.}~\bibnamefont{Onishchukov}},
  \bibnamefont{and} \bibinfo{author}{\bibfnamefont{U.}~\bibnamefont{Peschel}},
  \bibinfo{journal}{Phys. Rev. Lett.} \textbf{\bibinfo{volume}{109}},
  \bibinfo{pages}{093903} (\bibinfo{year}{2012}).

\bibitem[{\citenamefont{Dienst et~al.}(2013)\citenamefont{Dienst, Casandruc,
  Fausti, Zhang, Eckstein, Hoffmann, Khanna, Dean, Gensch, and
  Winnerl}}]{Dienst:2013gk}
\bibinfo{author}{\bibfnamefont{A.}~\bibnamefont{Dienst}},
  \bibinfo{author}{\bibfnamefont{E.}~\bibnamefont{Casandruc}},
  \bibinfo{author}{\bibfnamefont{D.}~\bibnamefont{Fausti}},
  \bibinfo{author}{\bibfnamefont{L.}~\bibnamefont{Zhang}},
  \bibinfo{author}{\bibfnamefont{M.}~\bibnamefont{Eckstein}},
  \bibinfo{author}{\bibfnamefont{M.}~\bibnamefont{Hoffmann}},
  \bibinfo{author}{\bibfnamefont{V.}~\bibnamefont{Khanna}},
  \bibinfo{author}{\bibfnamefont{N.}~\bibnamefont{Dean}},
  \bibinfo{author}{\bibfnamefont{M.}~\bibnamefont{Gensch}}, \bibnamefont{and}
  \bibinfo{author}{\bibfnamefont{S.}~\bibnamefont{Winnerl}},
  \bibinfo{journal}{nature materials} \textbf{\bibinfo{volume}{12}},
  \bibinfo{pages}{535} (\bibinfo{year}{2013}).

\bibitem[{\citenamefont{Dong et~al.}(2013)\citenamefont{Dong, Zhu, Zhang, and
  Malomed}}]{Dong:2013ky}
\bibinfo{author}{\bibfnamefont{G.}~\bibnamefont{Dong}},
  \bibinfo{author}{\bibfnamefont{J.}~\bibnamefont{Zhu}},
  \bibinfo{author}{\bibfnamefont{W.}~\bibnamefont{Zhang}}, \bibnamefont{and}
  \bibinfo{author}{\bibfnamefont{B.~A.} \bibnamefont{Malomed}},
  \bibinfo{journal}{Phys. Rev. Lett.} \textbf{\bibinfo{volume}{110}},
  \bibinfo{pages}{250401} (\bibinfo{year}{2013}).

\bibitem[{\citenamefont{Franzosi et~al.}(2011)\citenamefont{Franzosi, Livi,
  Oppo, and Politi}}]{Franzosi:2011ft}
\bibinfo{author}{\bibfnamefont{R.}~\bibnamefont{Franzosi}},
  \bibinfo{author}{\bibfnamefont{R.}~\bibnamefont{Livi}},
  \bibinfo{author}{\bibfnamefont{G.}~\bibnamefont{Oppo}}, \bibnamefont{and}
  \bibinfo{author}{\bibfnamefont{A.}~\bibnamefont{Politi}},
  \bibinfo{journal}{Nonlinearity} \textbf{\bibinfo{volume}{24}},
  \bibinfo{pages}{R89} (\bibinfo{year}{2011}).

\bibitem[{\citenamefont{Gericke et~al.}(2008)\citenamefont{Gericke, W{\"u}rtz,
  Reitz, Langen, and Ott}}]{Gericke:2008tq}
\bibinfo{author}{\bibfnamefont{T.}~\bibnamefont{Gericke}},
  \bibinfo{author}{\bibfnamefont{P.}~\bibnamefont{W{\"u}rtz}},
  \bibinfo{author}{\bibfnamefont{D.}~\bibnamefont{Reitz}},
  \bibinfo{author}{\bibfnamefont{T.}~\bibnamefont{Langen}}, \bibnamefont{and}
  \bibinfo{author}{\bibfnamefont{H.}~\bibnamefont{Ott}}, \bibinfo{journal}{Nat.
  Phys.} \textbf{\bibinfo{volume}{4}}, \bibinfo{pages}{949}
  (\bibinfo{year}{2008}).

\bibitem[{\citenamefont{Christodoulides and
  Eugenieva}(2001)}]{Christodoulides:2001tm}
\bibinfo{author}{\bibfnamefont{D.~N.} \bibnamefont{Christodoulides}}
  \bibnamefont{and} \bibinfo{author}{\bibfnamefont{E.~D.}
  \bibnamefont{Eugenieva}}, \bibinfo{journal}{Phys. Rev. Lett.}
  \textbf{\bibinfo{volume}{87}}, \bibinfo{pages}{233901}
  (\bibinfo{year}{2001}).

\bibitem[{\citenamefont{Trombettoni and
  Smerzi}(2001{\natexlab{b}})}]{Trombettoni:2001tb}
\bibinfo{author}{\bibfnamefont{A.}~\bibnamefont{Trombettoni}} \bibnamefont{and}
  \bibinfo{author}{\bibfnamefont{A.}~\bibnamefont{Smerzi}},
  \bibinfo{journal}{J. Phys. B} \textbf{\bibinfo{volume}{34}},
  \bibinfo{pages}{4711} (\bibinfo{year}{2001}{\natexlab{b}}).

\bibitem[{\citenamefont{Buonsante and Penna}(2008)}]{Buonsante:2008fe}
\bibinfo{author}{\bibfnamefont{P.}~\bibnamefont{Buonsante}} \bibnamefont{and}
  \bibinfo{author}{\bibfnamefont{V.}~\bibnamefont{Penna}}, \bibinfo{journal}{J.
  Phys. A} \textbf{\bibinfo{volume}{41}}, \bibinfo{pages}{175301}
  (\bibinfo{year}{2008}).

\bibitem[{\citenamefont{Oxtoby and Barashenkov}(2007)}]{Oxtoby:2007gq}
\bibinfo{author}{\bibfnamefont{O.}~\bibnamefont{Oxtoby}} \bibnamefont{and}
  \bibinfo{author}{\bibfnamefont{I.}~\bibnamefont{Barashenkov}},
  \bibinfo{journal}{Phys. Rev. E} \textbf{\bibinfo{volume}{76}},
  \bibinfo{pages}{036603} (\bibinfo{year}{2007}).

\bibitem[{\citenamefont{Hennig et~al.}(2010)\citenamefont{Hennig, Dorignac, and
  Campbell}}]{Hennig:2010gy}
\bibinfo{author}{\bibfnamefont{H.}~\bibnamefont{Hennig}},
  \bibinfo{author}{\bibfnamefont{J.}~\bibnamefont{Dorignac}}, \bibnamefont{and}
  \bibinfo{author}{\bibfnamefont{D.}~\bibnamefont{Campbell}},
  \bibinfo{journal}{Phys. Rev. A} \textbf{\bibinfo{volume}{82}},
  \bibinfo{pages}{053604} (\bibinfo{year}{2010}).

\bibitem[{\citenamefont{Aubry}(1997)}]{Aubry:1997wo}
\bibinfo{author}{\bibfnamefont{S.}~\bibnamefont{Aubry}},
  \bibinfo{journal}{Physica D} \textbf{\bibinfo{volume}{103}},
  \bibinfo{pages}{201} (\bibinfo{year}{1997}).

\bibitem[{\citenamefont{Marin and Aubry}(1996)}]{Marin:1996wv}
\bibinfo{author}{\bibfnamefont{J.}~\bibnamefont{Marin}} \bibnamefont{and}
  \bibinfo{author}{\bibfnamefont{S.}~\bibnamefont{Aubry}},
  \bibinfo{journal}{Nonlinearity} \textbf{\bibinfo{volume}{9}},
  \bibinfo{pages}{1501} (\bibinfo{year}{1996}).

\bibitem[{\citenamefont{Proville and Aubry}(1999)}]{Proville:1999vb}
\bibinfo{author}{\bibfnamefont{L.}~\bibnamefont{Proville}} \bibnamefont{and}
  \bibinfo{author}{\bibfnamefont{S.}~\bibnamefont{Aubry}},
  \bibinfo{journal}{Eur. Phys. J. B} \textbf{\bibinfo{volume}{11}},
  \bibinfo{pages}{41} (\bibinfo{year}{1999}).

\bibitem[{\citenamefont{Campbell et~al.}(1983)\citenamefont{Campbell,
  Schonfeld, and Wingate}}]{Campbell:1983ta}
\bibinfo{author}{\bibfnamefont{D.}~\bibnamefont{Campbell}},
  \bibinfo{author}{\bibfnamefont{J.}~\bibnamefont{Schonfeld}},
  \bibnamefont{and} \bibinfo{author}{\bibfnamefont{C.}~\bibnamefont{Wingate}},
  \bibinfo{journal}{Physica D} \textbf{\bibinfo{volume}{9}}, \bibinfo{pages}{1}
  (\bibinfo{year}{1983}).

\bibitem[{\citenamefont{Campbell}(1980)}]{Campbell:1980ua}
\bibinfo{author}{\bibfnamefont{D.~K.} \bibnamefont{Campbell}},
  \bibinfo{journal}{Annals of Physics} \textbf{\bibinfo{volume}{129}},
  \bibinfo{pages}{249} (\bibinfo{year}{1980}).

\bibitem[{\citenamefont{Rumpf}(2004)}]{Rumpf:2004en}
\bibinfo{author}{\bibfnamefont{B.}~\bibnamefont{Rumpf}},
  \bibinfo{journal}{Phys. Rev. E} \textbf{\bibinfo{volume}{70}},
  \bibinfo{pages}{016609} (\bibinfo{year}{2004}).

\bibitem[{\citenamefont{Kivshar and Campbell}(1993)}]{Kivshar:1993vf}
\bibinfo{author}{\bibfnamefont{Y.~S.} \bibnamefont{Kivshar}} \bibnamefont{and}
  \bibinfo{author}{\bibfnamefont{D.~K.} \bibnamefont{Campbell}},
  \bibinfo{journal}{Phys. Rev. E} \textbf{\bibinfo{volume}{48}},
  \bibinfo{pages}{3077} (\bibinfo{year}{1993}).

\bibitem[{\citenamefont{Ng et~al.}(2009)\citenamefont{Ng, Hennig, Fleischmann,
  Kottos, and Geisel}}]{Ng:2009tu}
\bibinfo{author}{\bibfnamefont{G.~S.} \bibnamefont{Ng}},
  \bibinfo{author}{\bibfnamefont{H.}~\bibnamefont{Hennig}},
  \bibinfo{author}{\bibfnamefont{R.}~\bibnamefont{Fleischmann}},
  \bibinfo{author}{\bibfnamefont{T.}~\bibnamefont{Kottos}}, \bibnamefont{and}
  \bibinfo{author}{\bibfnamefont{T.}~\bibnamefont{Geisel}},
  \bibinfo{journal}{New J. Phys.} \textbf{\bibinfo{volume}{11}},
  \bibinfo{pages}{073045} (\bibinfo{year}{2009}).

\end{thebibliography}

\newpage
%\newpage
\section{Supplemental Material}

\subsection{A cut through the dynamical phase diagram}
\begin{figure}[b]
    \centering
    \includegraphics[width=0.35\textwidth]{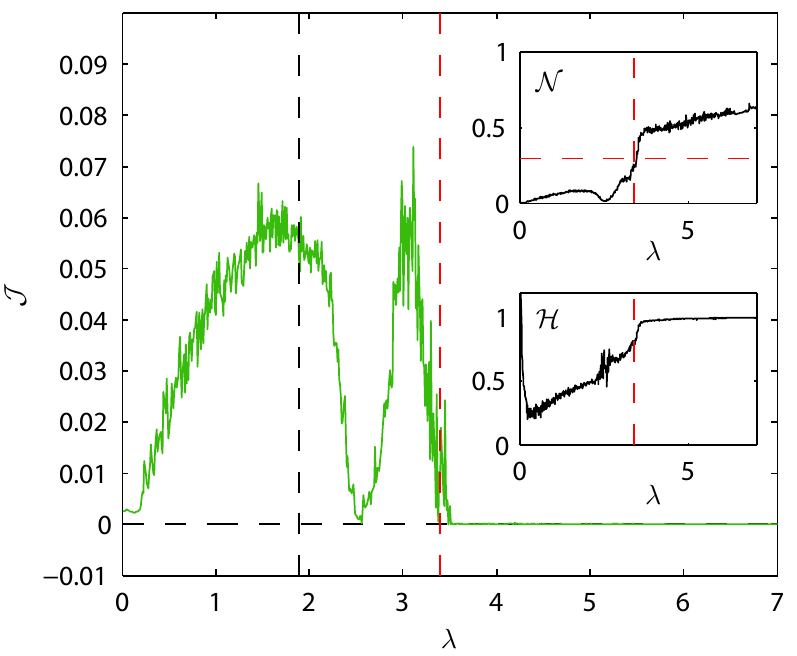}
    \caption{Cross-section through Fig.(\ref{fig:colorportrait_alpha1}) at $\cos(p_0)=0.88$. The black vertically dashed line corresponds to the predicted transition between the moving breather and the DB regime of \cite{Trombettoni:2001wl}, the red vertically dashed lines indicate our prediction Eq.(\ref{lambda_critical}). The horizontal red line in the $\mathcal{N}_{thrs}$- inset denotes the critical norm at the transition from our analytics in very good agreement with the numerical data. }
    \label{fig:crosss0}
\end{figure}

The dynamical phase diagram of Fig.~\ref{fig:colorportrait_alpha1} shows a remarkable transition from diffusive to moving breather behavior back to diffusive motion and than to self trapping for $\cos(p_0)\approxeq0.88$. In Fig.~\ref{fig:crosss0} we shows curves of the order parameters as a function of the nonlinearity $\lambda$ for $\cos(p_0)=0.88$, illustrating the sharp features in the order parameters at the transitions between the different regimes.

\subsection{Definition of the threshold $\mathcal{N}_{thrs}$} 
\begin{figure}[!t]
    \centering
    \includegraphics[width=0.5\textwidth]{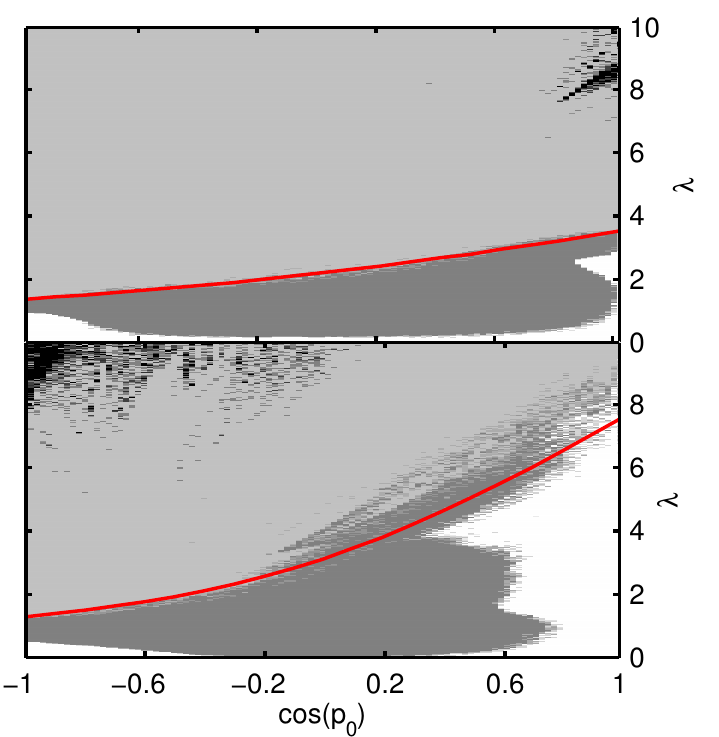}
    \caption{Dynamical phase diagram (left) $M=1001$ $\alpha_0=1$, (right) $M=1001$ $\alpha_0=4$ obtained with a pdf threshold of the maximum local norm  of $p_T=10^{-5}$.}
    \label{fig:phaseportrait_1001_10^-5}
\end{figure}

In order to find a suitable upper threshold of the maximal local norm in the diffusive regime we calculate the cumulative probability density function (pdf) of $\mathcal{N}$ in the diffusive case where the single site norms are known to be exponentially distributed $p=M\cdot e^{-Mx}$ \cite{Ng:2009tu}. Assuming statistical independents in this regime gives the pdf of the maximum single site norm as $p_{max}=M^2 (1-\exp(-Mx))^{M-1} \cdot \exp(-Mx)$. A DB however is localized over several sites around the maximum. Therefore we need to calculate the PDF of the local norm within a range of $a$ sites on either side of the maximum. The PDF of the sum of two single-site norms is given by the convolution of the two PDFs. For $r$ sites with exponentially distributed norms we can thus calculate the PDF of norms iteratively by $p_r=p_{r-1}*p$, where $*$ denotes convolution. The PDF of the local norm in the range of $2a$ sites around the maximum can consequently be expressed as $p_{\mathcal{N}}=p_{2a}*p_{max}$.  Let us examine an initial condition which leads to vanishing current $\mathcal{J}$. We will consider it to be self-trapped if its evolution leads to a maximum local norm $\mathcal{N}$ that is sufficiently unlikely to be found in the diffusive state. We define as the threshold $\mathcal{N}_{thrs}$ the norm for which the PDF  falls below $p_T=10^{-4}$. We find $\mathcal{N}_{thrs}=0.028$ for $\alpha_0=1$ and $\mathcal{N}_{thrs}=0.032$ for $\alpha_0=4$.
However, our results are not sensitive to the exact value of this parameter $p_T$. Assuming $p_T=10^{-5}$ yields $\mathcal{N}_{thrs}=0.031$ for $\alpha_0=1$ and $\mathcal{N}_{thrs}=0.035$ for $\alpha_0=4$ which result in the dynamical phase diagrams shown in Fig.~\ref{fig:phaseportrait_1001_10^-5}, which differ only minimally from those of Fig.~\ref{fig:colorportrait_alpha1}b. 

\subsection{Order parameters for $\alpha=4$}
\begin{figure}[!b]
      \centering 
      	\includegraphics[width=0.5\textwidth]{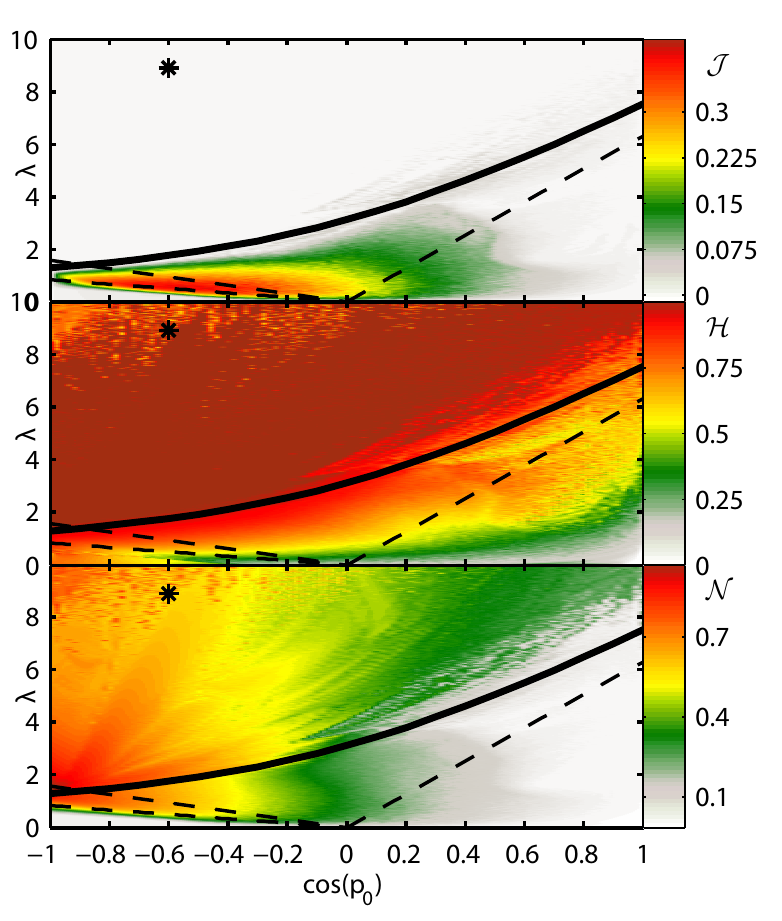}
      \caption{Colormap of the the three parameters $\mathcal{J}$ (top), $\mathcal{H}$ (top) and $\mathcal{N}$  (bottom) for $\sqrt{\alpha_0}=a=2$ and system size $M=1001$. The black dashed lines where predicted in~\cite{Trombettoni:2001wl} to mark the transition between different dynamical regimes. The solid line is our analytical estimate on the transition between the moving breather- and the self-trapping regime, Eq.(\ref{lambda_critical}).}
      	\label{fig:colorportrait_alpha4}
\end{figure}

Figure~\ref{fig:colorportrait_alpha4} shows the three order parameters $\mathcal{J}$, $\mathcal{H}$ and $\mathcal{N}$ for $\alpha=4$, which give rise to the dynamical phase diagram for $\alpha=4$ Fig.~\ref{fig:colorportrait_alpha1}b (bottom). 

\end{document}